\documentclass[aps,prl,preprint,superscriptaddress]{revtex4-1}
\usepackage{amsmath}
\usepackage{epstopdf}
\usepackage{graphicx}
\begin{document}

% Use the \preprint command to place your local institutional report
% number in the upper righthand corner of the title page in preprint mode.
% Multiple \preprint commands are allowed.
% Use the 'preprintnumbers' class option to override journal defaults
% to display numbers if necessary
%\preprint{}

%Title of paper
\title{Screening Charged Impurities and Lifting the Orbital Degeneracy in Graphene by Populating Landau Levels}

% repeat the \author .. \affiliation  etc. as needed
% \email, \thanks, \homepage, \altaffiliation all apply to the current
% author. Explanatory text should go in the []'s, actual e-mail
% address or url should go in the {}'s for \email and \homepage.
% Please use the appropriate macro foreach each type of information

% \affiliation command applies to all authors since the last
% \affiliation command. The \affiliation command should follow the
% other information
% \affiliation can be followed by \email, \homepage, \thanks as well.
\author{Adina Luican-Mayer}
\affiliation{Department of Physics and Astronomy, Rutgers University, Piscataway, NJ 08854, USA}

\author{Maxim Kharitonov}
\affiliation{Department of Physics and Astronomy, Rutgers University, Piscataway, NJ 08854, USA}

\author{Guohong Li}
\affiliation{Department of Physics and Astronomy, Rutgers University, Piscataway, NJ 08854, USA}

\author{ChihPin Lu}
\affiliation{Department of Physics and Astronomy, Rutgers University, Piscataway, NJ 08854, USA}

\author{Ivan Skachko}
\affiliation{Department of Physics and Astronomy, Rutgers University, Piscataway, NJ 08854, USA}

\author{Alem-Mar B. Goncalves}
\affiliation{Department of Physics and Astronomy, Rutgers University, Piscataway, NJ 08854, USA}

\author{K. Watanabe}
\affiliation{Advanced Materials Laboratory, National Institute for Materials Science, 1-1 Namiki, Tsukuba, 305-0044, Japan}

\author{T. Taniguchi}
\affiliation{Advanced Materials Laboratory, National Institute for Materials Science, 1-1 Namiki, Tsukuba, 305-0044, Japan}

\author{Eva Y. Andrei}
\affiliation{Department of Physics and Astronomy, Rutgers University, Piscataway, NJ 08854, USA}

%Collaboration name if desired (requires use of superscriptaddress
%option in \documentclass). \noaffiliation is required (may also be
%used with the \author command).
%\collaboration can be followed by \email, \homepage, \thanks as well.
%\collaboration{}
%\noaffiliation

\date{\today}

\begin{abstract}
We report the observation of an isolated charged impurity in graphene and present direct evidence of the close connection between the screening properties of a 2D electron system and the influence of the impurity on its electronic environment. Using scanning tunneling microscopy and Landau level spectroscopy we demonstrate that in the presence of a magnetic field the strength of the impurity can be tuned by controlling the occupation of Landau-level states  with a gate-voltage.   At low occupation the impurity is screened becoming essentially invisible. Screening diminishes as states are filled until, for fully occupied Landau-levels, the unscreened impurity significantly perturbs the spectrum in its vicinity. In this regime we report the first observation of Landau-level splitting into discrete states due to lifting the orbital degeneracy. 
\end{abstract}

% insert suggested PACS numbers in braces on next line
% insert suggested keywords - APS authors don't need to do this
%\keywords{}

%\maketitle must follow title, authors, abstract, \pacs, and \keywords
\maketitle

% body of paper here - Use proper section commands
% References should be done using the \cite, \ref, and \label commands

Charged-impurities are the primary source of disorder and scattering in two dimensional (2D) electron systems\cite{RevModPhys.54.437}.  They produce a spatially localized signature in the density of states (DOS) which, for impurities located at the surface, is readily observed with scanning tunneling microscopy and spectroscopy (STM/STS)\cite{mizes1989long}. In this respect graphene\cite{RevModPhys.81.109}\cite{abergel2010properties,*morg} \cite{Evareview}, \cite{RevModPhys.84.1067}, with its electronic states strictly at the surface, provides a unique playground for elucidating the role of impurities in 2D \cite{RevModPhys.84.1067} \cite{khveshchenko2006coulomb,*chen2008charged,*zhang2009origin,*wehling2010resonant,*PhysRevB.81.045409,*RevModPhys.83.407}\cite{pereira2007coulomb,*PhysRevLett.99.246802}. They are   particularly important in the presence of a magnetic field when the quantization of the 2D electronic spectrum into highly degenerate Landau levels (LL) gives  rise to the  quantum Hall effect (QHE).  In this regime charged-impurities are expected to lift the orbital-degeneracy causing each LL in their immediate vicinity to split into discrete sub-levels \cite{PhysRevB.83.235104,*PhysRevB.85.165423}. Thus far however the sub-levels were not experimentally accessible due to the difficulty to attain sufficiently clean samples that would allow isolating a single impurity. Instead, previous experiments \cite{yacoby1999electrical,*PhysRevLett.101.256802},\cite{yoshioka2006local,*PhysRevLett.102.026803,*miller2010real, *morgenstern2012scanning}    presented a picture of “bent” levels which  could be interpreted in a semi-classical framework in terms of electronic drift trajectories  moving along the equipotential lines defined by a dense distribution of charged-impurities.

In this work we employed high quality gated graphene devices which provided access to the electronic spectrum in the QHE regime in the presence of an isolated charged-impurity. We demonstrate that the strength of the impurity, as measured by its effect on the spectrum, can be controlled by tuning the LL occupation with a back-gate-voltage.  For almost empty LLs the impurity is screened and essentially invisible whereas at full LL occupancy screening is very weak and the impurity attains  maximum strength.  In the unscreened regime we experimentally resolve the underlying discrete quantum-mechanical spectrum arising from lifting the orbital-degeneracy.

The low energy spectrum of pristine graphene, consisting of two electron-hole symmetric Dirac cones, gives rise to a linear DOS, which vanishes at the charge neutrality point (CNP). In the presence of a magnetic field B, the spectrum is quantized into a sequence of LLs characteristic of massless Dirac fermions:
\begin{equation}
E_N=\pm\frac{v_F}{l_B}\sqrt{2|N|} ,N=0,\pm{1},\pm{2},... \label{LLi}
\end{equation}

where $v_F$ is the Fermi velocity, $ l_B=\sqrt{\hbar/eB}$ the magnetic length, e the electron charge, $\hbar$ the reduced Planck constant and $\pm$ refers to electron (hole) states with LL index $N>1 (N<1)$.   We employed LL spectroscopy \cite{Evareview}  to study the electronic properties of graphene and their modification in the presence of a charged-impurity. The LL spectra were obtained by measuring the bias voltage dependence of the differential tunneling conductance, $dI/dV$, which is proportional to the DOS, $DOS(E,\bf{r})$ , at the tip position $\bf{r}$. Here $V=(E-E_F)/e$  is the bias voltage and E the energy measured relative to the Fermi level, $E_F$.

\begin{figure}
\includegraphics [width=\textwidth]{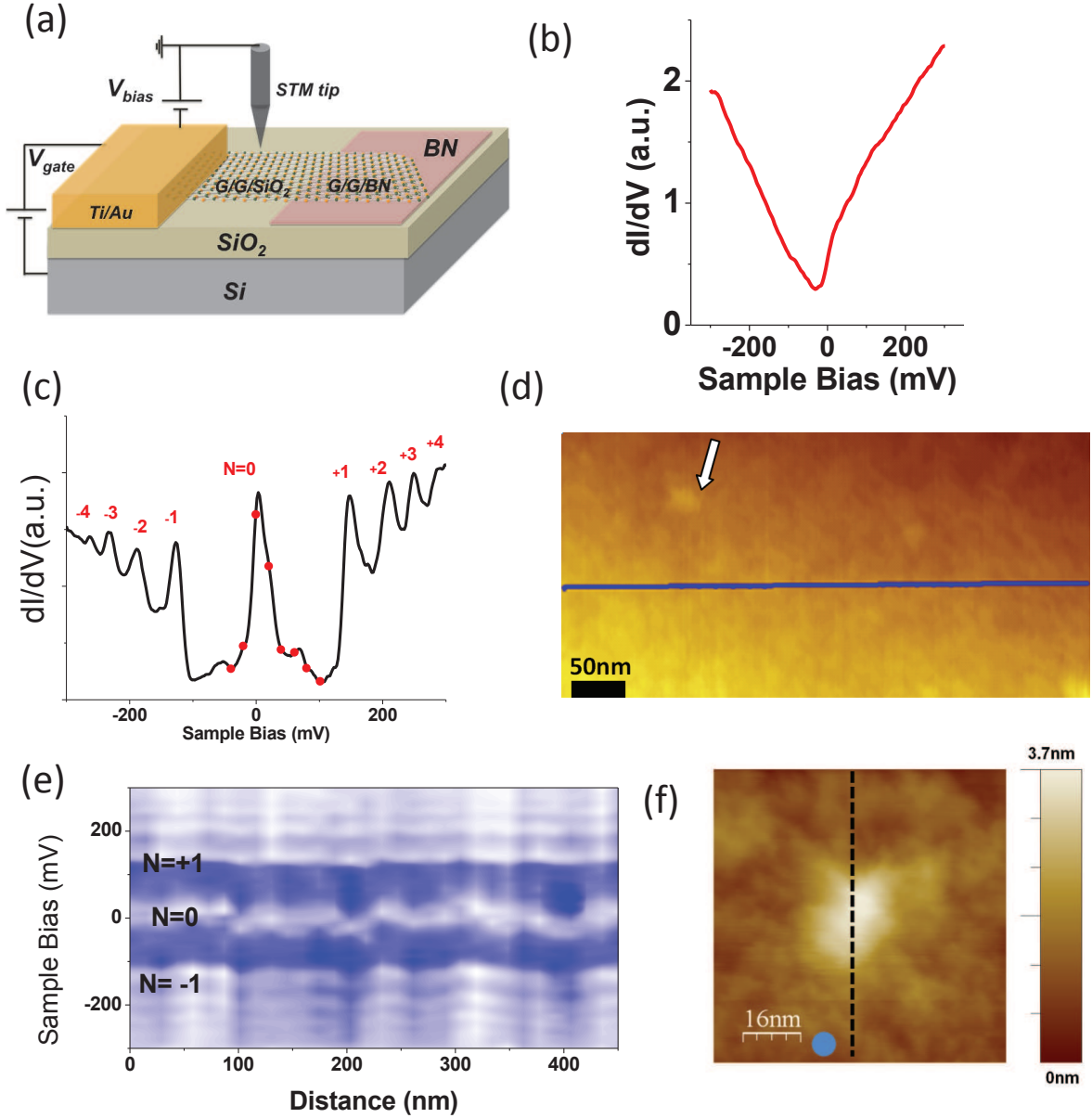}
 \caption{\label{fig1}(a) Schematics of gated graphene device illustrating the regions where the two stacked graphene layers are deposited directly on SiO$_2$  and on a flake of   h-BN in  (  $G/G/SiO_2$  and $G/G/BN$ ). The two graphene layers share the same electrode and  Fermi level.  (b) Zero field STS taken far from the impurity shown in panel d. (c) STS at  $B = 10 T$ and $V_g =0V$ shows well resolved quantized LLs. Red circles indicate  bias voltages of the maps in Figure \ref{fig2}.  (d) Large area topographic image indicating the line where the spectra in panel e were taken and the position of the isolated impurity. (e) STS line cut along the line shown in panel d; LLs with indices $ N=0$,  are clearly resolved. (f) STM topography zoom-into the area with an  isolated impurity $(V_B=250mV, I_t=20pA)$. The spectra in panels b, c and the LL map in Figure \ref{fig3} were taken at the position indicated by the dot. }
\end{figure}

Samples were prepared by exfoliating graphene from analyzer-grade HOPG and deposited on a doped Si back-gate capped with 300nm of chlorinated SiO$_2$\cite{PhysRevB.83.041405}.  In order to achieve high quality we used two superposed graphene layers twisted away from the standard Bernal stacking by a large angle. The large twist angle ensures that the spectrum of single layer graphene is preserved \cite{meyer2007structure,*li2009observation,*PhysRevLett.106.126802} \footnote{As shown by G. Li et al. Nat. Phys. 6,109, (2009), a  twist  between superposed  graphene layers gives rise to two peaks in the density of states (Van-Hove singularities) which flank the charge neutrality point and are separated from each other by an energy which increases with twist-angle.   For twist-angles exceeding 10 degrees  the low energy spectrum $(<1eV)$ is indistinguishable from that of  single layer graphene. The absence of Van-Hove singularities and  the single layer LL spectrum in the data reported here  provide direct evidence of layer decoupling. Although there is no topographic signature of the associated Moire pattern, which would require a very sharp tip, the above signatures are  taken as evidence for a large twist angle.} while reducing the random potential fluctuations due to substrate imperfections.Hexagonal boron-nitride (h-BN) flakes  which significantly reduce the corrugation of graphene\cite{xue2011scanning} were also employed  (Figure \ref{fig1}a), but the data reported is restricted to the SiO$_2$ substrate.

\begin{figure}
\includegraphics [width=\textwidth]{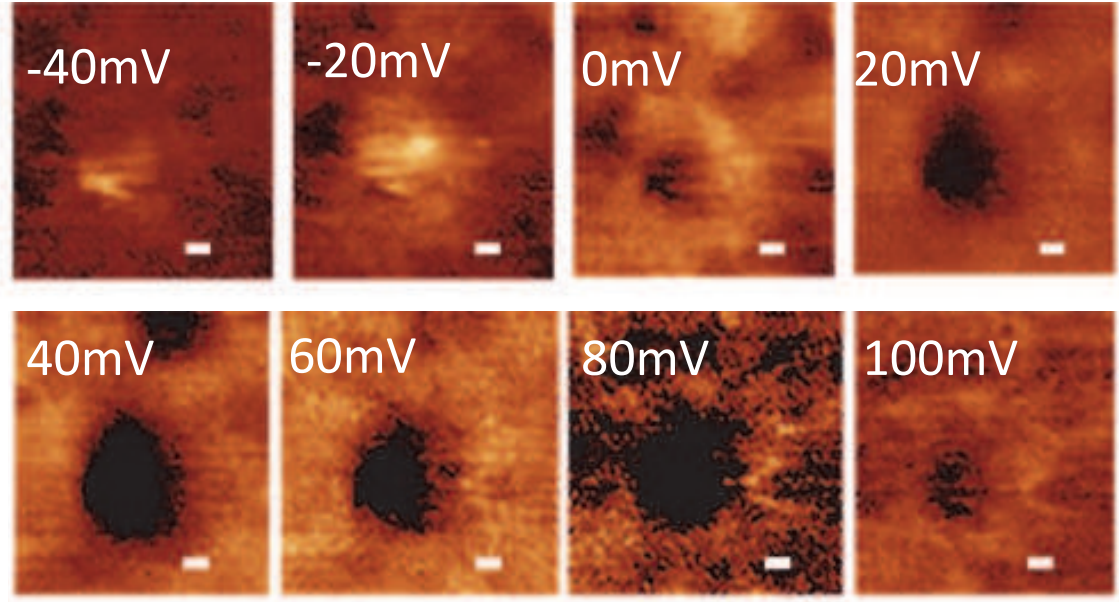}
 \caption{\label{fig2}Spatial $dI/dV$ maps at  $B=10T$ near the impurity taken at indicated bias voltages. Scale bar for all maps: $ 8.2nm= l_B$  .}
\end{figure}

Using the STM tip as a capacitive antenna \cite{li073701} we located  the samples at low temperature and performed spectroscopy measurements to identify  areas of interest.   A typical zero-field spectrum taken far from an impurity in Figure \ref{fig1}b reveals the V-shaped DOS characteristic of single layer graphene (SLG).  In finite-field the spectrum develops pronounced peaks (Figure \ref{fig1}c)  at energies corresponding to the LLs \cite{li2007observation,*PhysRevLett.102.176804} that are well resolved up to $N = 4$ in both electron and hole sectors, attesting to good sample quality. Fitting the field and level index dependence to Equation \ref{LLi} confirms the massless Dirac fermion nature of the quasiparticles with $v_F = 1.2 \times 10^6 m/s$, a value consistent with measurements on SLG. Charged-impurities were located using LL-spectroscopy to measure the separation between $E_F$ and the CNP which coincides with the $N=0$ level. LL-spectroscopy is more sensitive to the position of the CNP than the broad zero-field spectrum. The search for impurities starts with a topography image (Figure \ref{fig1}d) followed by STS within this area. An intensity map of the LLs as a function of position (Figure \ref{fig1}e) shows the fluctuations of the $N = 0$ level in response to charged-impurities. We focus on an area with a minimal number of impurities indicated by the arrow. Zooming into this area the impurity appears as an isolated bright region in the center of Figure \ref{fig1}f. To visualize its effect on the spatial distribution of the electronic-wavefunction  we measured constant energy DOS maps (Figure \ref{fig2}). The maps are roughly radially symmetric, consistent with a charged-impurity at the center. We note that for energies within a gap between LLs the electronic DOS (bright region) is tightly localized on the impurity. In contrast, for energies in the center of the LL the electronic DOS extends across the entire field of view while avoiding the impurity.  This fully supports the picture which attributes the QHE plateaus to the existence of localized impurity states in the gaps between LLs.

\begin{figure}

\includegraphics [width=\textwidth]{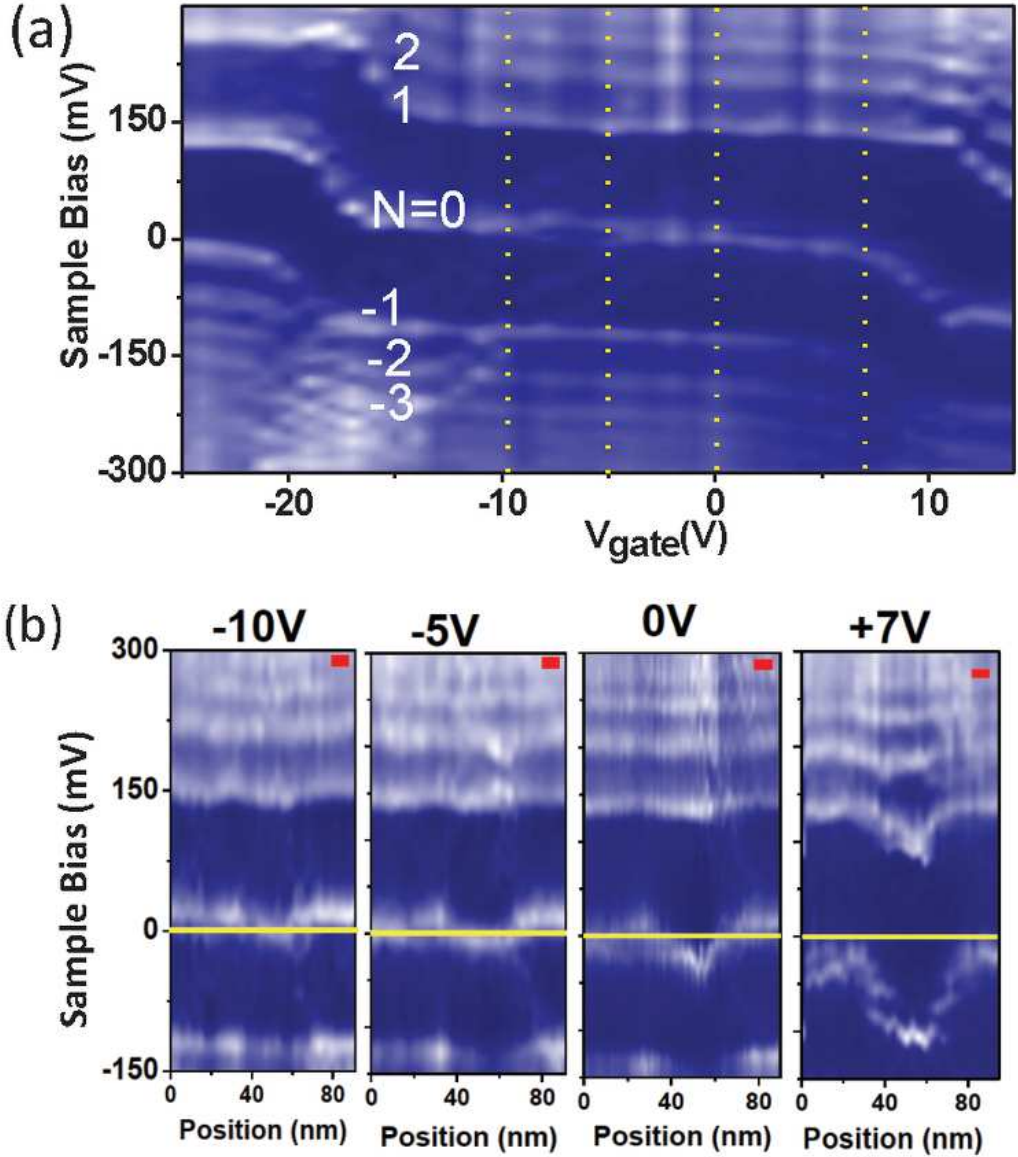}
 \caption{\label{fig3} Impurity screening by populating Landau levels. (a) DOS map at $B=10T$ showing evolution of LLs as a function of gate-voltage taken far from the impurity at the position indicated  in Figure \ref{fig1}f. Dashed lines indicate  gate-voltages at which the spectra in b were taken. (b) LL maps across the impurity for indicated gate-voltages. The distortion of the LL sequence by the impurity is strongest for filled levels $(V_g =+7V)$, diminishing as filling is reduced and becoming almost invisible at $V_g = -10V $.}
\end{figure}

We next studied the effect of LL occupancy (filling) by tuning the gate-voltage, $V_g$ , to progressively fill the LLs. The LL filling factor is $\nu=n/n_0(B)$ where $n\approx 7\times 10^{10}  V_g  (V) cm^{-2}$ is the carrier density and $n_0(B) =g_lg_vg_s\frac{Be}{2\pi\hbar}$   is the degeneracy/area of the LL.  Here $g_l=g_v=g_s=2$   represent the layer, valley and spin degeneracy respectively. Placing the STM tip far from the impurity we find that as $V_g$ is swept the LL peaks produce a distinctive step-like pattern seen as bright traces in the intensity map of Figure \ref{fig3}a \cite{dial2007high,PhysRevB.83.041405}. Each step consists of a nearly horizontal plateau separated from its neighbors by steep slopes. The separation between the centers of  steep segments, $\Delta V_g \approx 28V$, gives the LL degeneracy $\approx 2\times10^{12} cm^{-2}$  for $B=10T$, as expected for this double-layer sample \cite{PhysRevLett.107.216602,*PhysRevLett.108.076601}.  The plateau indicates that the Fermi-energy remains pinned within a narrow energy band around the center of the LL until the plateau states are filled. A further increase in $V_g$  populates the sparse states in the gap producing the steep slopes\footnote{ Although STM explores only a small area of the sample, the gate-voltage dependence of the data in Figure \ref{fig3}a reflects the available states in the entire sample including those that are outside the field of view of the STM. This is because the gate covers the entire sample and can populate all available states.}.

To explore the influence of the impurity on the LLs we follow the spatial evolution of spectra along a trajectory traversing it (Figure \ref{fig1}f) for a series of gate-voltages.   As shown in Figure \ref{fig3}b, for certain gate-voltages  the spectra become significantly distorted close to the impurity, with the $N=0$ level (and to a lesser extent higher order levels) shifting downwards toward negative energies. The downshift indicates an attractive potential produced by a positively charged-impurity. Its strength, as measured by the distortion of the $N=0$ LL, reveals a surprisingly strong dependence on LL filling.  In the range of gate-voltages corresponding to filling the $N =0$ LL  $(-15V < V_g < 9 V)$ the distortion grows monotonically with filling.   At small filling the distortion is almost absent indicating that the impurity is effectively   screened and it reaches its maximum value close to full occupancy. At full occupancy  the  $N=0$ level shifts  by as much as  $\approx 0.1 eV$ indicating that the effect would survive at room temperature. We note that this spectral distortion is only present in the immediate vicinity of the impurity.  Farther away no distortion is observed for all the carrier densities studied here.

We attribute the variation of the impurity strength with filling to the screening properties of the electron system. For a positively (negatively) charged-impurity and almost empty (full) LLs, unoccupied states necessary for virtual electron transitions are readily available in the vicinity of the impurity, resulting in substantial screening. By contrast for almost filled (empty)  LLs, unoccupied  states are scarce, which renders local screening inefficient.   These properties are readily understood by examining the local DOS, $D_s(E,r)=\int{d^2rD(E,r)/S}$, averaged over a finite-size region, S, around the impurity. Unlike the DOS averaged over the whole sample, $D_s(E,r)$, is manifestly particle-hole asymmetric within a given LL which translates to the particle-hole asymmetry of the local screening.

\begin{figure}
\includegraphics [width=0.8\textwidth]{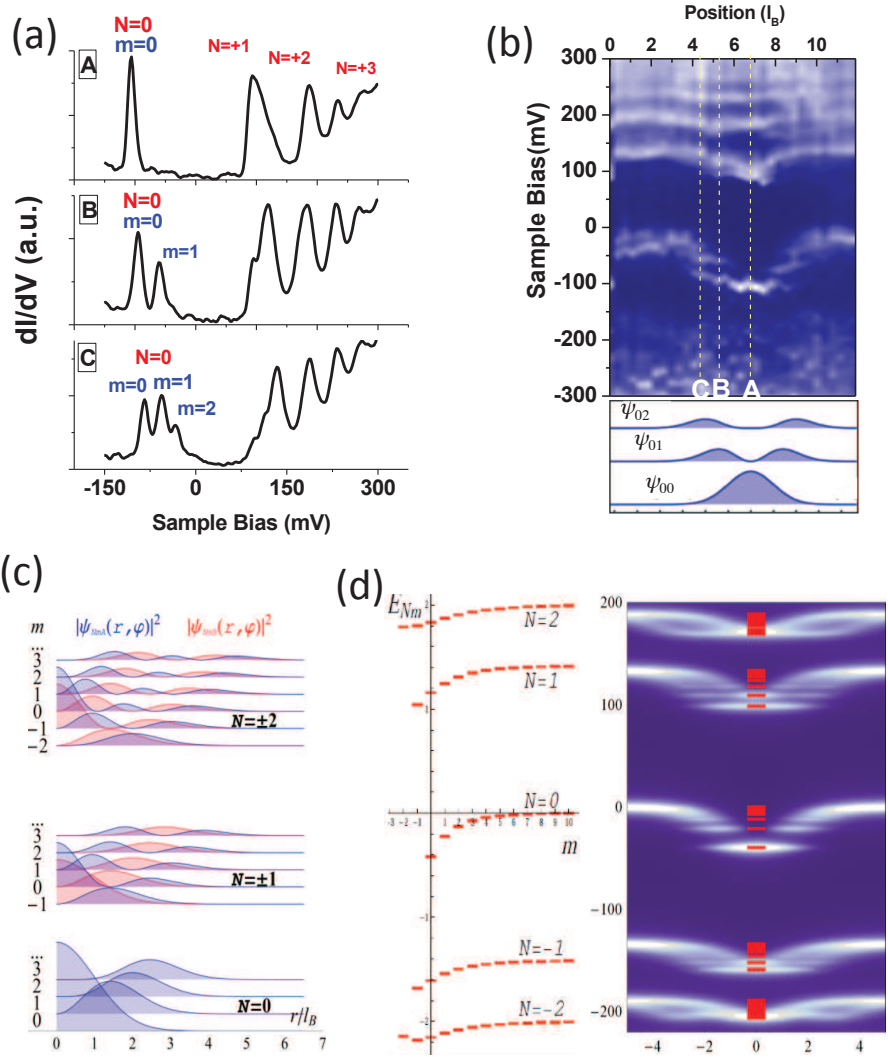}
 \caption{\label{fig4} Lifting the orbital-degeneracy. (a) $dI/dV$ spectra for $B=10T$ and $V_g=7V$ at the positions indicated in panel b reveal the appearance of peaks corresponding to states with $N=0$ and $m=0$, $m=0,1$ and $m=0,1,2$ as marked. (b). Top: LL map across the impurity for $V_g = 7V$. Dashed lines at distances $0nm, 13nm$ and $20nm$ from the center of the impurity indicate the position of the spectra in panel a. Bottom: calculated probability densities for states $\psi_{0m}$ with $m=0,1,2$  are consistent with the spatial distribution of the discrete spectral lines in the top panel. (c) Calculated probability densities on the two graphene sublattice A (blue) and B (red).  (d) Left panel: simulated spectrum near an impurity illustrating lifting the orbital-degeneracy in different LLs.  Right panel: simulated DOS near an impurity. Linewidth $\gamma= 0.05 v_F/l_B$ . Red lines represent the calculated energies, $E_{Nm}$ shown in the left panel.}
\end{figure}

Remarkably, when screening is minimal $(V_g =+7V)$  the $N = 0$ LL does not shift smoothly, but rather splits into a series of well resolved discrete spectral lines in the immediate vicinity of the impurity. As shown in Figure \ref{fig4}a,b the evolution of  the spectra radially outwards from the center of the impurity exhibits a progression  of peaks within the $N=0$ LL.  Starting with a single peak at the center of the impurity, it evolves into a well resolved double peak and then a triplet at distances $\approx 13 nm, 20 nm$ from the center respectively. This behavior can be understood by considering the quantum-mechanical electron motion in the presence of a magnetic-field and a charged-impurity. In one valley and for each spin projection, the two-component wave-function $\psi=(\psi_A,\psi_B )^T$ satisfies an effective Dirac Hamiltonian:

\begin{equation}
    \hat{H}\psi =E \psi, \hat{H}= \hat{H}_0+U(r),\hat{H}_0=\hbar v_F \boldsymbol{\sigma}(\bf{p}-e\bf{A}) \label{Dirac}
\end{equation}

Here, $\hat{H}_0$ corresponds to the case without the impurity, $\boldsymbol{\sigma} =(\sigma_x,\sigma_y)$ are the Pauli matrices in the sublattice space, $\bf{p}=-i\hbar\bf{\nabla}, \bf{B} = \bf{\nabla}\times{\bf{A}}$ , and $\bf{A}$ is the vector potential. We assume a radially symmetric impurity-potential $U(r)$, and neglect the Zeeman-effect. In the symmetric-gauge  $\bf{A}=\frac{1}{2} [\bf{B}\times \bf{r}]$  the eigenstates are characterized by the orbital quantum-number $m$. 
Solving  $\hat{H}_0$ yields the unperturbed spectrum in Equation \ref{LLi}, and the eigenfunctions $\psi^0_{NmA}(r),\psi^0_{NmB}(r)$  ( Figure \ref{fig4}c) where $m\ge -\mid{N}\mid$ . Since $E_N$  are independent of $m$, the LLs have infinite orbital-degeneracy. The impurity lifts this orbital-degeneracy and the eigenenergies split  into series of sublevels, $E_{Nm}$.

To illustrate impurity-induced orbital-splitting we numerically solved the problem for a Coulomb potential $U (r) =\frac{Z}{\kappa}\frac{e^2}{4\pi\epsilon_0}\frac{1}{\sqrt{r^2+a^2}}$ corresponding to a charge $Z$ located a distance $a$ below the graphene plane with $\kappa$ the effective dielectric constant and $\epsilon_0$ the permitivity of free-space. The resulting simulated-spectrum in the left panel of Figure \ref{fig4}d, shows that the orbital-degeneracy is lifted resulting in an $m$ dependent energy downshift.  The downshift is largest for $E_{00}$, and diminishes with increasing $m$ and/or $N$ where the unperturbed  LLs are approached.   For comparison with the STS data we calculated the local tunneling DOS  assuming a finite linewidth $\gamma$ :
     \begin{equation}
\label{eq3}
  D(E,\textbf{r})= 4\sum_{Nmi} \delta_\gamma(E-E_{Nm}) \psi^\dagger_{Nm}(\textbf{r}) \psi_{Nm}(\textbf{r})
     \end{equation}
Here $i$ is the sublattice index and  $\delta_\gamma(E-E_{Nm})=\gamma/[\pi((E-E_{Nm})^2+\gamma^2)]$ represents the broadened LL. The peak intensity is determined by the probability-density $ \psi^\dagger_{Nm}(\textbf{r}) \psi_{Nm}(\textbf{r})$ and is position dependent.  
 If $\gamma < \Delta{E_{Nm}}$ ($\Delta E_{Nm}$  spacing between adjacent levels) the discreteness of the spectrum is resolved, but for $\gamma \ge \Delta{E_{Nm}}$ ($\Delta E_{Nm}$   peaks of adjacent states overlap and merge into a continuous band.

Thus, even if the spectrum is discrete, but the resolution insufficient or if impurities are too close to each other, the measured $D(E,\bf{r})$  will still display  "bent" LLs, whose energies seemingly adjust to the local potential. The resulting simulated $D(E,\bf{r})$ , shown in Figure \ref{fig4}d (right-panel),   captures the main features of  the data. In particular,  upon approaching the impurity the $N = 0$ LL splits into well resolved discrete peaks, attributed to specific orbital states. In both experiment and simulation the states  $\psi_{0m}(\bf{r})$ with $m = 0,1,2$ are well resolved close to the impurity but higher order states, are less affected and their contributions to $D(E,\bf{r})$  merge into a continuous line. Similarly, the discreteness of the spectrum is not resolved for $N\not=0$, consistent with the weaker impurity effect at larger distances.  We note that for partial filling ($ V_g = -5V, 0V $)   as screening becomes more efficient and orbital-splitting is no longer observed the unresolved sublevels merge into continuous lines of "bent" Landau levels (Figure  \ref{fig3}b). Thus the capability to tune the strength of the impurity potential by the gate-voltage allows us to trace the evolution between the discrete and the previously observed quasi-continuous regimes \cite{yoshioka2006local,*PhysRevLett.102.026803,*miller2010real, *morgenstern2012scanning}.

We now turn to the effective charge of the impurity and how it’s screening is affected by LL occupancy. Although we do not have in-situ chemical characterization of the impurity we can assume based on chemical-analysis of similar samples that the most likely candidate consistent with our observations is an  $Na^+$ ion adsorbed on the surface of the SiO$_2$ substrate  \cite{constant2000deposition}. It is well known that $Na^+$  ions are ubiquitous in cleanrooms and laboratory environments and they are readily adsorbed on SiO$_2$ \cite{fripiat1971chimie}. Levels of $Na^+$  contamination as high as $10^{12} cm^{-2}$ can be reached within just a few days of exposure to human activity. For a $Na^+$  ion adsorbed on the substrate underneath the first graphene layer  $a\approx 0.6nm \ll  l_B$.  A rough estimate of its effect on the spectrum can be obtained by equating the measured energy shift $\Delta E_{00} \approx 0.1eV $, obtained at $V_g = +7V$ to the calculated value in first order perturbation theory:

$\Delta E_{00}=\frac{Z}{\kappa}\frac{e^2}{4\pi\epsilon_0}\langle \psi_{00}\mid \frac{1}{\sqrt{r^2+a^2}}\mid\psi_{00}\rangle=\frac{Z}{\kappa}\frac{e^2}{4\pi\epsilon_0 l_B}(\pi/2)^{(1/2)} (1-Erf(a/l_B))$

where $Erf(x)_{x\ll1}\to 0$ is the error function. Using $\Delta E_{00}\approx\frac{Z}{\kappa}\frac{e^2}{4\pi \epsilon_0 l_B}(\pi/2)^{1/2}= 0.1eV$ we obtain $Z/\kappa=2.5$ .  Factoring out the contribution of the $SiO_2$ substrate, $\kappa_{SiO_2}=4$, from the expression for the effective dielectric constant \cite{hwang} $\kappa=\kappa_{gr}(\kappa_{SiO_{2}}+1)/2$, gives $Z/\kappa_{gr}\cong1$ where $\kappa_{gr}$ is the static dielectric constant of graphene. This indicates that graphene provides no screening for gate-voltages corresponding to fully occupied LLs and that bare charge of the impurity is $ Z=+1$. In the limit of almost empty LLs ($V_g = -10V$) the estimated  $\kappa_{gr}\approx 5$ indicates that graphene strongly screens the impurity potential \cite{*[{This value is comparable to the  zero field RPA estimate from: }][{  for double layer graphene $\kappa_{gr}\approx 2.4$,$\kappa_{gr}=1+g_l g_s g_v \pi r_s/8\approx3.75$  suggesting that when the LLs are almost empty  screening of positive charges in graphene  is not very different from the  zero field case.  Here $r_s=4\pi e^2/h v_F (\kappa_{SiO_{2}}+1)$  is the dimensionless Wigner-Seitz radius which measures the relative strength of the potential and kinetic energies in an interacting quantum Coulomb system with linear dispersion. We note that for single layer graphene, $g_l=1$, screening would be significantly weaker, $\kappa_gr\approx2.4$. }] PhysRevB.75.205418}. The absence of screening for filled states implies strong Coulomb interactions when $E_F$ lies  in a gap consistent with the observation of a fractional QHE in suspended graphene \cite{du2009fractional,*bolotin2009observation} .   

This work demonstrates that screening in graphene is controlled by LL occupancy and that   it is possible to tune charge-impurities and their effect on the environment by applying a gate voltage or by varying the magnetic field. In the limit of strong screening, corresponding to low LL occupancy   a positive impurity is essentially invisible, while for high LL occupancy it is unscreened and lifts the orbital-degeneracy of the LLs in its vicinity.  Due to the large enhancement of the effective fine-structure constant in graphene a charged-impurity with $Z\ge1$  is expected to become  supercritical\cite{RevModPhys.81.109}\cite{pereira2007coulomb,*PhysRevLett.99.246802} but tuning the effective charge to observe supercritically is extremely difficult\cite{wang2012mapping,*wang2013observing}. The ability demonstrated here to tune the strength of the impurity in-situ opens the door to exploring Coulomb criticality  and to investigate a hitherto inaccessible regime of criticality in the presence of a magnetic field \cite{PhysRevB.83.235104,*PhysRevB.85.165423}.

\begin{acknowledgments}
Funding was provided by DOE-FG02-99ER45742 (E.Y.A and G.L), Lucent (A.L-M), NSF DMR 1207108 (I.S., C.P.L) , DOE DE-FG02-99ER45790 (M. K.), Brazilian agency Capes BEX 5115-09-4 ( A.M.B.G). We wish to thank M. Aronson for the monochromator grade HOPG crystal.
\end{acknowledgments}

% Create the reference section using BibTeX:
%

\end{document}